\documentclass[letterpaper, 10 pt, conference]{ieeeconf}  % Comment this line out if you need a4paper

\IEEEoverridecommandlockouts                              % This command is only needed if 
\newenvironment{IEEEkeywords}{\begin{center}\bfseries Keywords\end{center}\begin{quote}}{\end{quote}}

\usepackage[utf8]{inputenc}
\usepackage{amsmath} % assumes amsmath package installed
\DeclareUnicodeCharacter{202F}{\,} % map narrow no-break space to thin space
\usepackage{cite}
\usepackage{graphicx}     % already in most IEEE templates
\usepackage{multirow}
\usepackage{booktabs}
\usepackage{url}
\usepackage{makecell}
\usepackage{balance} 
\usepackage{hyperref}

\title{\LARGE \bf
Pre-Clinical Latency Characterization of VRxBioRelax: A Real-Time EMG Biofeedback System for Muscle Relaxation in Virtual Reality
}

\author{
Melanie Baumgartner$^{1,*}$, 
Raphael Weibel$^{1,*}$, 
Tobias Hoesli$^{1,*}$, 
Aydin Javadov$^{1}$, 
Rayna Ney$^{1}$,\\ 
Helen Schwerdt$^{2}$, 
Florian von Wangenheim$^{1}$, 
Joseph Ollier$^{1}$%
\thanks{*Equal contribution.}%
\thanks{$^{1}$Mobiliar Lab for Analytics, ETH Zurich, Switzerland \newline
{\tt\small \{mbaumgart, rweibel, thoesli, ajavadov, rayney, fwangenheim, jollier\}@ethz.ch}}%
\thanks{$^{2}$ HESAV School of Health Science, HES-SO, Switzerland\newline
{\tt\small helen.schwerdt@hesav.ch}}%
}

\begin{document}

\maketitle
\thispagestyle{empty}
\pagestyle{empty}

\begin{abstract}
Chronic tension in the upper trapezius (UT), often caused by poor ergonomics, prolonged posture, or psychological stress, contributes to musculoskeletal discomfort, headaches, and impaired interoceptive awareness. Although surface electromyography (sEMG) biofeedback can promote UT relaxation, traditional systems using conventional displays often fail to sustain engagement. Virtual reality (VR) offers a more immersive alternative, provided that latency remains below perceptual thresholds. We introduce \textit{VRxBioRelax}, a closed-loop VR biofeedback system that streams sEMG data from Delsys Trigno\textsuperscript{\textregistered} Avanti sensors via MQTT to a Unity scene. Muscle activation drives a dynamic dawn-to-dusk landscape synchronized with a progressive muscle relaxation protocol. To validate system responsiveness, 87,716 EMG samples from the NinaPro DB2 dataset were replayed at $\sim$75~Hz. Timestamps at four key stages—acquisition, Root Mean Square (RMS) processing, network receipt, and rendering—revealed mean latencies of 0.50~ms (processing), 5.62~ms (network), and 19.22~ms (rendering), yielding an average end-to-end delay of 25.34~ms. Notably, 99.3\% of frames arrived within 50~ms. One-sided \textit{t}-tests confirmed mean latency was significantly lower than both the 30~ms VR comfort limit ($t_{87\,715}=-25.2$, $p=5.9{\times}10^{-140}$) and the 50~ms clinical benchmark ($t_{87\,715}=-133.3$, $p<10^{-300}$). These findings support \textit{VRxBioRelax} for use in remote interoceptive training, stress reduction, and telepresence-enabled rehabilitation
\end{abstract}

%%%%%%%%%%%%%%%%%%%%%%%%%%%%%%%%%%%%%%%%%%%%%%%%%%%%%%%%%%%%%%%%%%%%%%%%%%%%%%%%


\section{INTRODUCTION}
Chronic tension in the upper trapezius (UT), a muscle spanning from neck to shoulder, is widespread in stagnant, tech-heavy lifestyles \cite{LeVasseur2022,Patselas2021}. Prolonged computer use and poor ergonomics overload the UT, causing tightness, headaches, and altered shoulder mechanics \cite{Javed2024,Li2024}. Beyond mechanical stress, high levels of UT activation measured by electromyography (EMG) also reflect emotional stress and reduced interoceptive awareness \cite{Wijsman2013}. Progressive Muscle Relaxation (PMR) techniques, stretching, trigger-point release, or guided tension--release can relieve tension and improve well-being \cite{Ahmed2023_Feasibility,Ahmed2024_StressMeditation,Saleh2020}. Traditional surface-EMG biofeedback, usually visualized on flat screens, teaches users to modulate muscle activity, but lacks sustained engagement \cite{Blume2020,Toepp2023}. Virtual reality (VR) shows promise in enhancing immersion and motivation: For example, Pardini et al. \cite{Pardini2023} reported higher engagement with web-based VR PMR (without real-time feedback), while Weibel et al. \cite{Weibel2023} found VR-based HRV biofeedback boosts parasympathetic tone. However, to date, no current system delivers real-time, muscle-specific EMG feedback within an embodied VR PMR experience.

\begin{figure}[t]
  \centering
  \begin{minipage}[b]{.45\linewidth}
    \centering
    \includegraphics[width=\linewidth]{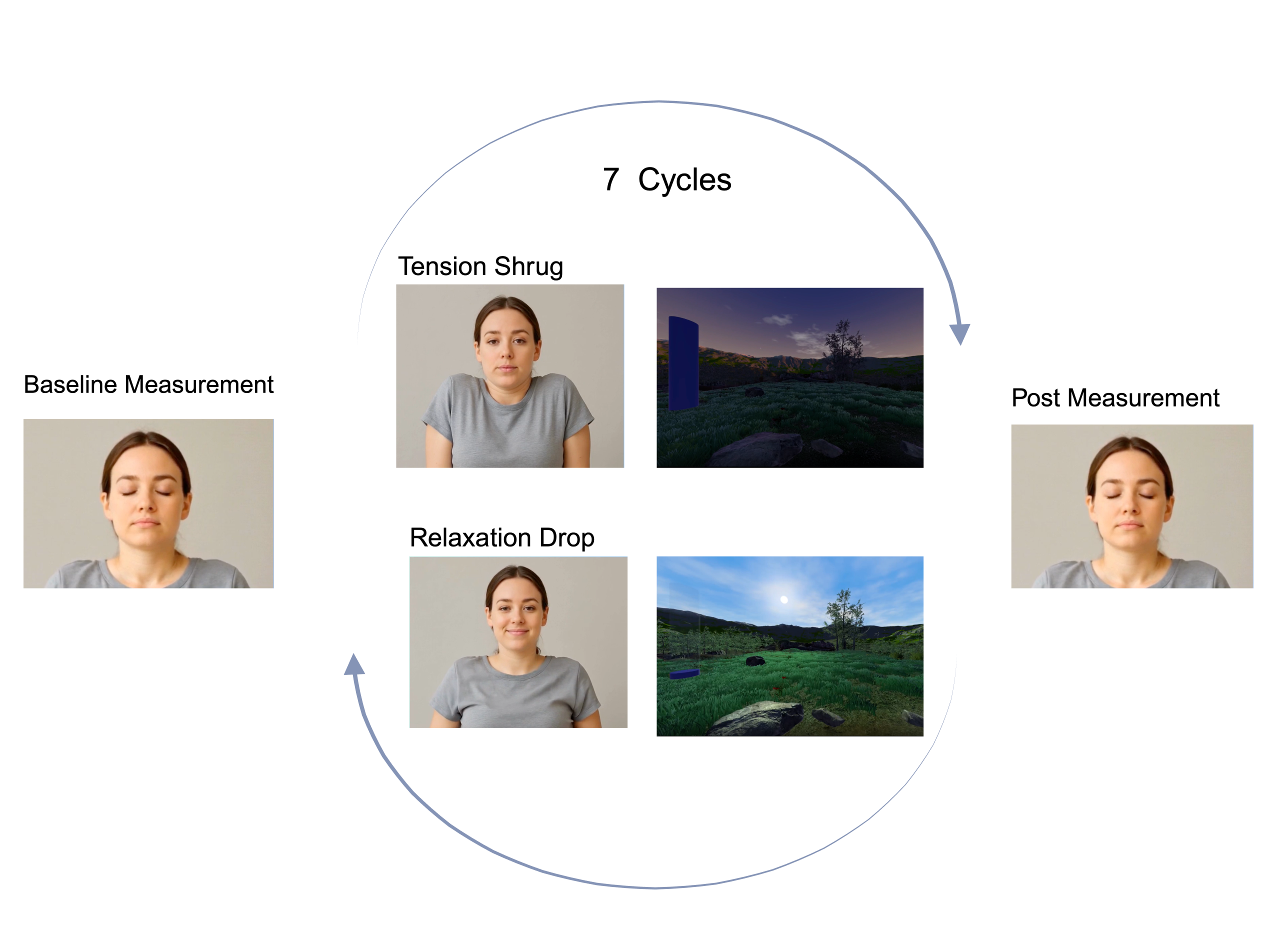}
    
    {\small (a)}
  \end{minipage}\hfill
  \begin{minipage}[b]{.45\linewidth}
    \centering
    \includegraphics[width=\linewidth]{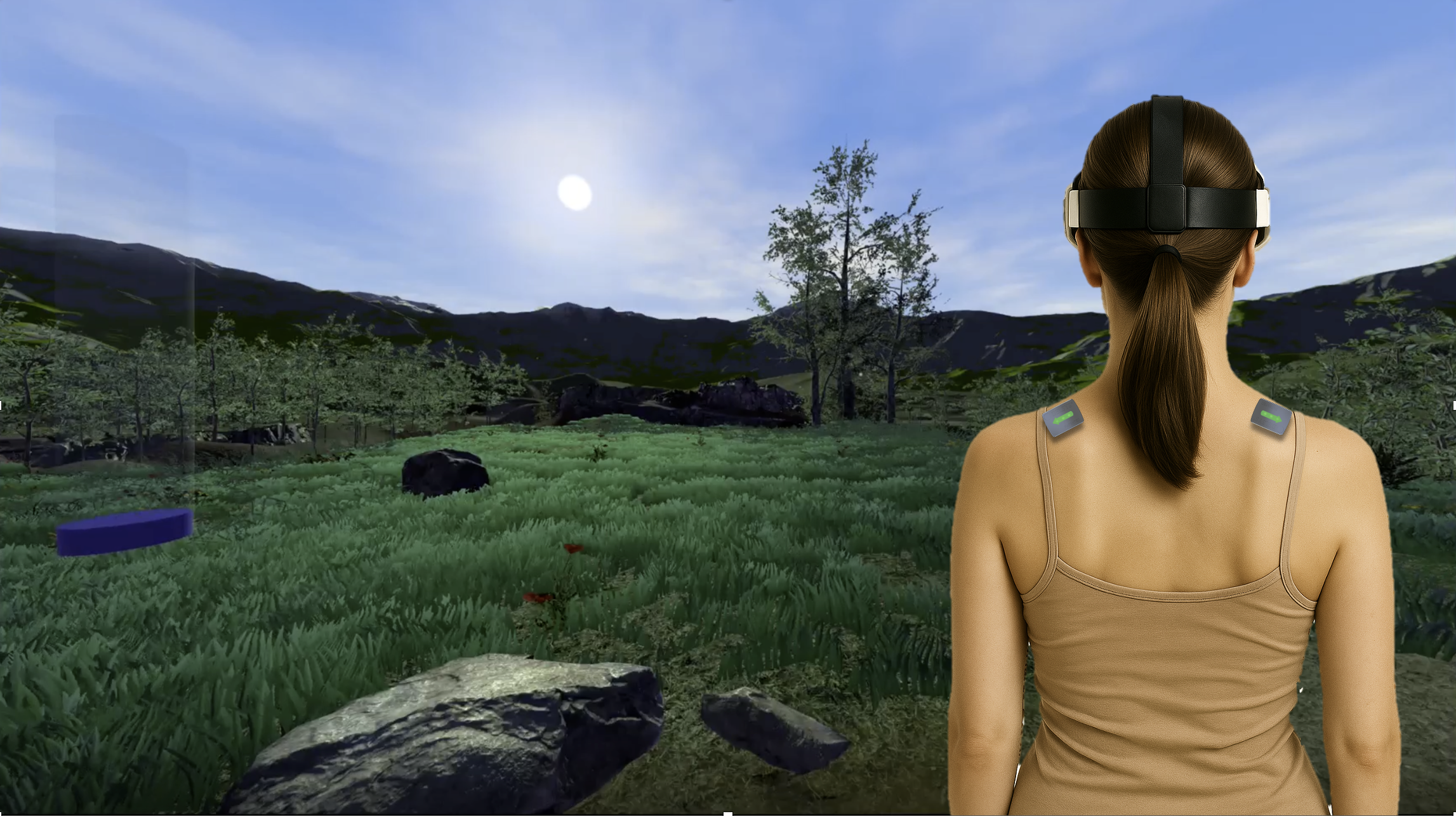}
    
    {\small (b)}
  \end{minipage}
  \caption{Illustration of the VRxBioRelax biofeedback mechanism: (a) EMG signal processing in real-time; (b) corresponding modulation of the immersive VR environment.}
  \label{fig:biofeedback-vr}
\end{figure}

We present \textit{VRxBioRelax}, a VR biofeedback system that captures UT EMG using Delsys Trigno\textsuperscript{\textregistered} Avanti sensors, transmits data via MQTT to a Unity-based VR app, and as shown in Fig.~\ref{fig:biofeedback-vr}, the application couples real‑time signal processing (a) with a dynamic dawn–dusk scene (b).The goal of our application is to deepen interoceptive awareness, support sustained engagement, and offer timely, physiologically relevant feedback for stress reduction.

\subsection{The Critical Role of Technical Validation in Therapeutic Technologies}
Before testing \textit{VRxBioRelax} in human trials, it is critical to confirm the real-time reliability of the system. Latency, even at tens of milliseconds, can disrupt the perception-action feedback loop, reducing the effectiveness of biofeedback or VR therapeutic applications \cite{Albert2017, Farago2023,Fornes2025,Safavynia2013,Toepp2023}. In this study, we conducted a full-stack latency assessment covering (raw) EMG acquisition, streaming, and VR rendering to ensure delays stay well below perceptual thresholds to enhance user experience. 

\subsection{Minimising Latency to Enhance Immersion and Sensorimotor Control}
Effective VR-based biofeedback relies on a tight, real-time coupling between a user’s bodily activation and the corresponding sensory outcome\cite{Ahmed2024_StressMeditation,Fornes2025}. Even modest delays can decouple and disrupt intended effort from visual reward, breaking the associative loop that drives motor relearning. Whereas entertainment-focused VR can tolerate latencies up to 70 ms for foveated rendering \cite{Albert2017} or even 150 ms in non-reflexive titles \cite{Fornes2025}, therapeutic feedback must be substantially tighter. Prior work in EMG biofeedback demonstrated that 64 ms windows with 6.25 Hz updates strike a practical balance between responsiveness and stability \cite{Toepp2023}, but emerging evidence suggests that delays above 50 ms begin to blunt sensorimotor adaptation \cite{Hadjiosif2025}. Likewise, end-to-end latencies beyond $\sim$40 ms are subjectively perceived as “laggy” by even casual users \cite{Farago2023,Fornes2025}, although some rehabilitation games may feel acceptable up to 90 ms \cite[19]{Kantan2022}. Together, these findings argue for a general latency ceiling of 50 ms below both the objective drop in learning rate and the subjective perceptual threshold. To provide a further safety margin against any residual delay in the EMG signal’s electromechanical transduction, we set an ambitious design target of 30 ms end-to-end latency in VRxBioRelax. This threshold preserves the sense of agency of the users, drives the feedback effectively 'instantaneously', and maximises both immersion and the therapeutic efficacy of our activation and relaxation biofeedback signals. 

% \noindent\textbf{Study Aims—} Specifically, we examined whether (i) the mean end-to-end latency falls below \textbf{30 ms} and (ii) at least \textbf{98 \%} of feedback frames are delivered within \textbf{50 ms}.

\subsection{Aim of This Study}
Prior to human-subject deployment, this study aims to validate \textit{VRxBioRelax}'s real-time performance by testing whether:
\begin{enumerate}
    \item The mean end-to-end latency will be significantly below 30 ms.
    \item At least 98\% of all feedback frames (out of 87\,716 samples) will arrive within 50\,ms (i.e., fewer than 2\% exceed that threshold).
\end{enumerate}

% Requires: \usepackage{booktabs, multirow}
% \begin{table*}[ht]
%     \centering
%     \begin{tabular}{@{}l rrrrrrr @{}}
%         \toprule
%         \multirow{2}{*}{\textbf{Latency Stage}} & \multicolumn{7}{c}{\textbf{Latency (ms)}} \\
%         \cmidrule(l){2-8}
%          & Mean & SD & Min & 25th pct & Median & 75th pct & Max \\
%         \midrule
%         Processing   & 0.50 & 0.00 & 0.50 & 0.50 & 0.50 & 0.50 & 0.50 \\
%         Network      & 5.62 & 3.42 & -1.68 & 2.98 & 5.40 & 8.00 & 28.05 \\
%         Rendering    & 19.22 & 54.86 & 0.00 & 10.00 & 15.00 & 20.00 & 1194.00 \\
%         End-to-End   & 25.34 & 54.79 & 7.35 & 15.76 & 20.50 & 27.39 & 1197.83 \\
%         \bottomrule
%     \end{tabular}
%     \caption{Summary statistics of latency stages (in milliseconds) based on a sample of 87,716 observations.}
%     \label{tab:latency_summary}
% \end{table*}

\section{METHODS}
\subsection{Dataset Selection and Preparation}

\begin{table*}[!t]
\centering
\caption{Latency descriptors for each pipeline stage ($n = 87\,716$; all times in ms).
For the end-to-end loop the mean delay is {\bfseries 24.7 ms} below the 50 ms clinical
target and {\bfseries 4.7 ms} below the stricter 30 ms VR-comfort limit (one-sided
$t$-tests: $p<10^{-300}$ and $p=5.9{\times}10^{-140}$, respectively).}
\label{tab:latency_compact}
\renewcommand{\arraystretch}{1.12}
\begin{tabular}{@{}lcccc@{}}
\toprule
\textbf{Stage} &
\textbf{Mean $\pm$ SD} &
\textbf{Median [IQR]} &
\textbf{$P_{95}$} &
\textbf{$t$-test vs target} \\[1pt]
\midrule
Processing & 0.50\,$\pm$\,0.00 & 0.50 [0.50–0.50] & 0.50 & — \\
Network    & 5.62\,$\pm$\,3.42 & 5.40 [2.98–8.00] & 11.34 & — \\
Rendering  & 19.2\,$\pm$\,54.9 & 15.0 [10–20]     & 28.0  & — \\
\midrule[0.5pt]
End-to-end & 25.3\,$\pm$\,54.8 & 20.5 [15.8–27.4] & 32.3 &
\makecell[l]{%
50 ms: $\Delta=-24.7$, $t=-133.3$ ${}^{***}$\\
30 ms: $\Delta=-\,4.7$, $t=-25.2$ ${}^{***}$} \\
\bottomrule
\end{tabular}
\end{table*}

We validated our EMG-to-VR pipeline using publicly available surface EMG recordings from the NinaPro DB2 dataset \cite{Atzori2014}, which employs the same Delsys Trigno Avanti sensor system used in our application. A single 5-minute trial ($N = 87\,716$ samples) was selected to match the duration of our planned progressive muscle relaxation exercise. Over a 5~min recording we collected 87,716 samples, thus an effective rate of $\approx$292~Hz for our Root Mean Square (RMS)-streaming pipeline. Raw EMG signals were first converted to their RMS envelope in Python and down-sampled to approximately 75~Hz to reflect the real-time streaming rate of our system. 
% ==== PLACEHOLDER FOR YOUR ADDITIONAL FIGURE ====
\begin{figure}[!t]
  \centering
  \includegraphics[width=\linewidth]{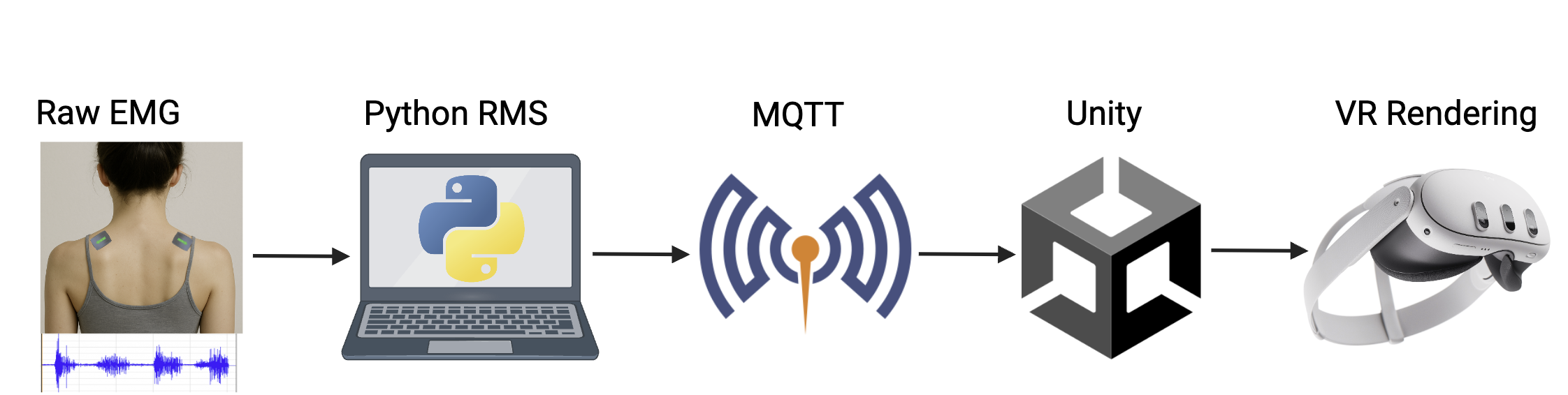}
  \caption{Raw electromyographic (EMG) signals from the upper trapezius are captured using Delsys Trigno® Avanti sensors and streamed to a custom Python interface that computes real-time root mean square (RMS) amplitude. The processed data are transmitted via MQTT protocol to a Unity-based virtual reality (VR) environment. This environment dynamically maps EMG activity to changes in visual elements and is rendered through a VR headset, enabling closed-loop, immersive biofeedback for muscle tension regulation.}
  \label{fig:method-workflow}
\end{figure}
% ===============================================
\subsection{Data Replay, Streaming, and Latency Analysis}
To characterize end-to-end latency in the \textit{VRxBioRelax} feedback pipeline, we implemented a Python-based EMG replay client that streams the NinaPro DB2 EMG-Dataset using the Paho–MQTT library. Each transmitted JSON packet includes a sequential identifier, the original sensor timestamp, and the timestamp corresponding to RMS computation. On the receiving end, a Unity 6 LTS application—developed using the \textit{M2MqttUnity} framework—subscribes to the MQTT topic and logs packet arrival times at three critical stages: (1) upon reception, (2) immediately before the visual update loop, and (3) after rendering the 3D feedback element. All measurements were conducted with the virtual environment rendered on a Meta Quest 3 headset. Both the MQTT broker (Mosquitto v2.0) and the Unity application were executed locally on a Lenovo ThinkPad T14 Gen 4 laptop (AMD Ryzen 7 PRO 7840U, Radeon 780M Graphics, 8 cores / 16 threads, 32~GB RAM) running Microsoft Windows 11 Education (Build 26100). The Meta Quest 3 and host PC were wirelessly connected via a shared 5~GHz Wi-Fi 6 mobile hotspot provided by an Android smartphone, ensuring low-latency communication across the local network. Offline, logs from the two systems are merged on the packet identifier, and four latency metrics are derived by simple differencing: processing (sensor → computation), network (computation → receipt), rendering (receipt → visual update), and end-to-end (sensor → final display). This modular pipeline yields fine-grained insight into performance across acquisition, transport, and visualization stages (see Fig.~\ref{fig:method-workflow}).\footnote{Complete source code for the replay client and latency-analysis scripts is available at \url{https://osf.io/zhmtg/files/osfstorage} (accessed 3 July 2025).}

\section{RESULTS}

\begin{figure}[t]
  \centering
  \includegraphics[width=\linewidth]{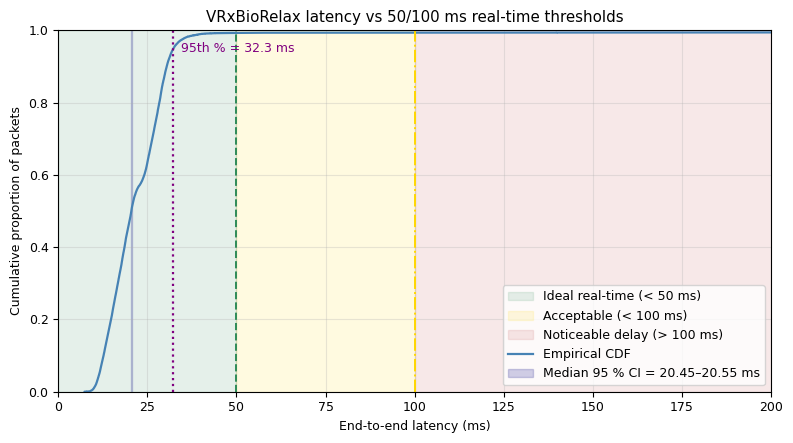}
  \caption{%
    Empirical cumulative distribution of end-to-end latency
    (\(n=87{,}716\) packets). Shaded bands mark evidence-based limits for closed-loop human–machine interaction: latencies below
    \(50\,\mathrm{ms}\) are virtually imperceptible for
    sensorimotor control, whereas delays up to
    \(100\,\mathrm{ms}\) remain acceptable but can already
    attenuate motor accuracy~\cite{Hadjiosif2025}.
    The navy band shows the bootstrap $95\%$ confidence
    interval of the sample median
    (20.45–20.55~ms) and the dotted line indicates the
    95th percentile (32~ms).  Consequently, \(\sim99\%\) of
    packets fall inside the “ideal” $<50$~ms window,
    outperforming the \(\approx90\)~ms loop delay~\cite{Kantan2022}.}
  \label{fig:latencycdf}
\end{figure}

\begin{figure}[ht]
  \centering
  \includegraphics[width=0.85\linewidth]{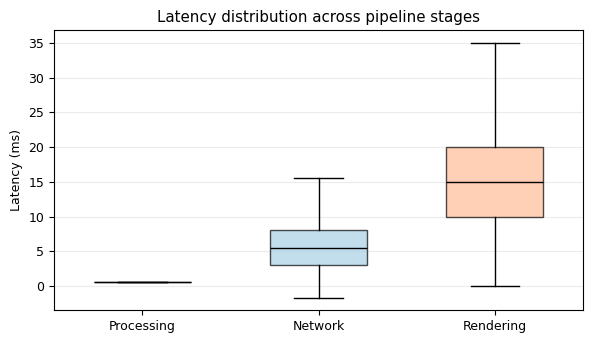}
  \caption{Latency distributions across the three pipeline stages: Processing -- box collapsed at 0.50 ms, showing constant compute time; Network -- median 5.4 ms, IQR (3.0--8.0 ms), full range $-1.7$--15.5 ms; Rendering -- median 15.0 ms, IQR (10.0--20.0 ms), full range 0--35.0 ms. Whiskers span the minimum and maximum observed values.}
  \label{fig:latency-boxplot}
\end{figure}

\begin{figure}[!htb]
    \centering
    \includegraphics[width=0.85\linewidth]{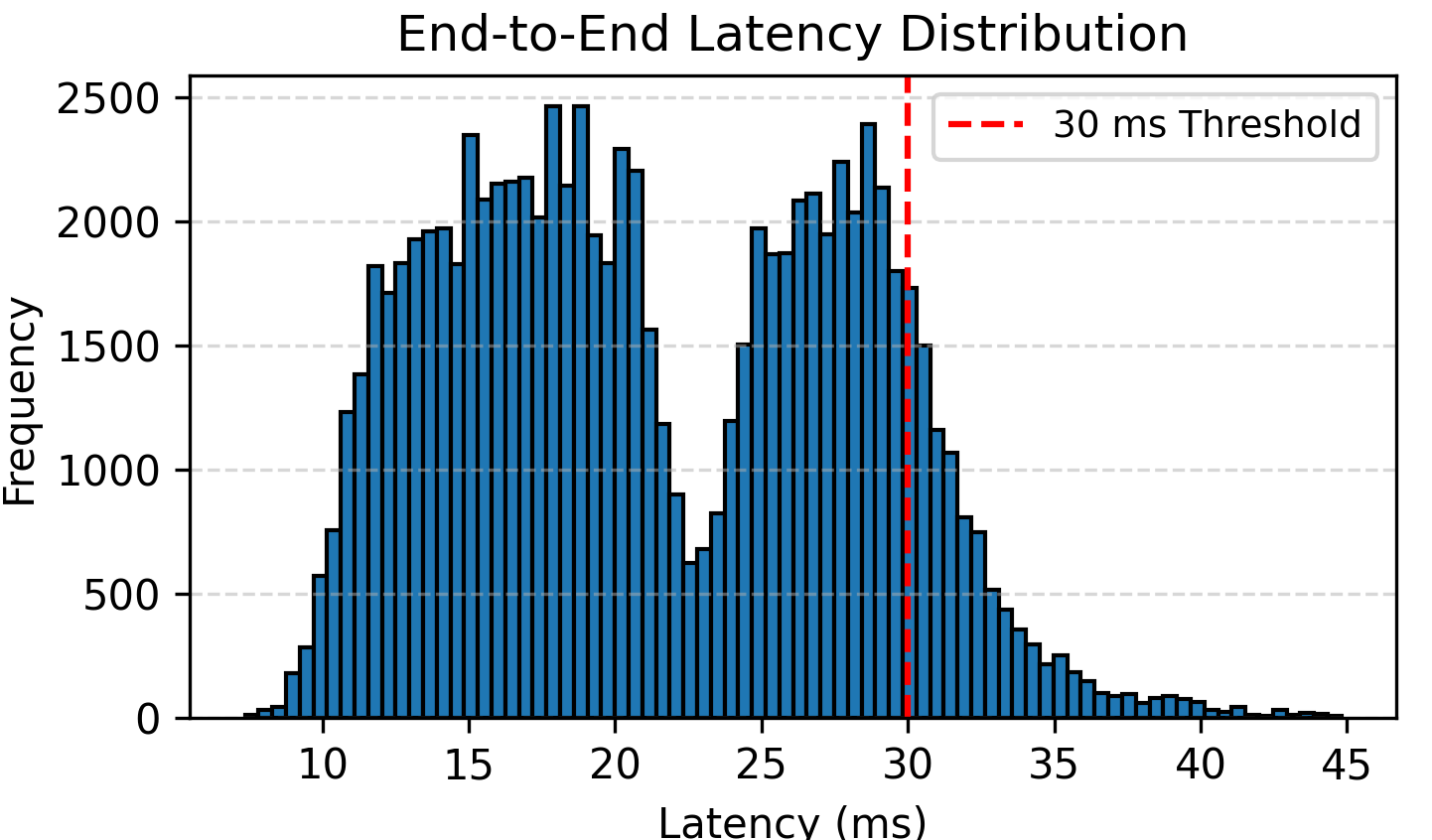}
    \caption{Histogram of end-to-end latency (1 ms bins; axis truncated at
45 ms, covering 99.7 \% of packets).\,
The dashed red line marks the 30 ms VR-comfort threshold.
The sample mean (25.3 ms) lies well to the left and is
significantly lower than \emph{both} the 30 ms and the clinical
50 ms targets (\emph{cf.} Table~\ref{tab:latency_compact},
one-sided $t$-tests).}
    \label{fig:latency-histogram}
\end{figure}

Processing latency remained essentially constant at 0.5~ms, as shown in Table~\ref{tab:latency_compact}, reflecting the minimal computational cost of RMS calculation. Network latency averaged 5.62~ms (median 5.40~ms), with 75\% of packets arriving within 8~ms; a small negative minimum value (--1.68~ms) was attributed to minor clock synchronization drift  (see Fig.~\ref{fig:latency-boxplot}). Rendering latency averaged 19.2~ms (median 15~ms), approximately one frame at 60~Hz, with occasional outliers reaching up to 1.2~seconds, likely due to rare frame stalls or garbage collection pauses. Overall, the end-to-end latency (from EMG acquisition to VR rendering) was approximately 25~ms on average (median 20.5~ms), remaining well below the 50~ms threshold generally considered imperceptible in real-time VR biofeedback applications. The cumulative distribution is plotted in Fig.~\ref{fig:latencycdf}.

While over 98\% of latency values clustered tightly around the median (network $\approx$ 5~ms, render $\approx$ 15~ms, end-to-end $\approx$ 20~ms), a small number of outliers (1--2\%) showed spikes between 50--120~ms. These rare delays were traced to expected system-level behaviors: OS thread interruptions, Unity garbage collection, and brief MQTT callback congestion. Despite creating a long-tailed distribution, these outliers were infrequent and did not compromise real-time performance.

Only 0.8\% of frames exceeded 45~ms. A one-sample \textit{t}-test against the 30~ms target yielded $t(87\,715) = -25.2$, $p < 0.001$, confirming the mean latency is significantly below 30~ms. A Wilcoxon signed-rank test likewise gave $p < 0.001$. These results demonstrate our pipeline consistently operates well under the 30~ms threshold.

\section{DISCUSSION}
This pre-clinical evaluation confirms that the \textit{VRxBioRelax} system reliably delivers real-time EMG biofeedback within a virtual reality environment, adhering to stringent latency and signal quality standards. With a median end-to-end latency of 20.5~ms and 98\% of frames processed under 45~ms, the system operates well below commonly accepted perceptual thresholds for interactive biofeedback and VR applications. These results provide strong evidence that our implementation is technically suitable for interoceptive training, muscle relaxation, and future therapeutic use. The low processing and network latencies (median $\approx$0.5~ms and $\approx$5.4~ms, respectively) demonstrate efficient data handling and minimal overhead in transmission. Rendering latency averaged 19.2~ms, close to the frame time of 60~Hz displays ($\approx$16.7~ms), suggesting the visual pipeline is fast enough to maintain congruency between muscle state and VR feedback. Although rare outliers (1--2\%) exceeded 50~ms---due to expected system-level factors such as OS thread preemption or Unity garbage collection---these did not significantly impair the overall responsiveness of the system. Despite these encouraging results, some limitations should be acknowledged. First, this evaluation used simulated EMG playback rather than live recordings from active participants, and thus may not capture noise or artifact issues that arise during actual muscle activity. Second, our benchmarks used a single configuration: Meta Quest 3 ←→ Lenovo ThinkPad T14 Gen 4 laptop linked via the same 5 GHz Wi-Fi 6 hotspot on a smartphone, which may not generalise to slower CPUs, wired or enterprise networks, or other HMDs.
%Second, our tests ran on a single hardware configuration; variations in CPU performance, network infrastructure, or VR headset capabilities could affect latency profiles in other settings. 
To address these gaps, future work will involve live-subject trials to assess system responsiveness under natural movement and artifact conditions, as well as cross-platform benchmarking across diverse hardware and network environments. We will also explore adaptive buffering and prediction algorithms to mitigate occasional outliers caused by OS preemption or garbage collection. By extending validation to real users and varied deployments, we aim to establish VRxBioRelax as a robust and scalable tool for clinical biofeedback. 

\section{CONCLUSION}
Latency benchmarks show that VRxBioRelax operates well below perceptual thresholds, enabling reliable, immersive biofeedback for UT relaxation. These guarantees justify the planned clinical trial and support subsequent extension to at-home stress-management protocols.

\addtolength{\textheight}{-12cm}   % This command serves to balance the column lengths

%%%%%%%%%%%%%%%%%%%%%%%%%%%%%%%%%%%%%%%%%%%%%%%%%%%%%%%%%%%%%%%%%%%%%%%%%%%%%%%%

%%%%%%%%%%%%%%%%%%%%%%%%%%%%%%%%%%%%%%%%%%%%%%%%%%%%%%%%%%%%%%%%%%%%%%%%%%%%%%%%

%%%%%%%%%%%%%%%%%%%%%%%%%%%%%%%%%%%%%%%%%%%%%%%%%%%%%%%%%%%%%%%%%%%%%%%%%%%%%%%%
\section*{APPENDIX}
\textbf{Code availability.} All source code used for the development and implementation of the closed-loop sEMG-VR biofeedback system is publicly available on the Open Science Framework (OSF): \href{https://osf.io/zhmtg/files/osfstorage}{OSF project files}.

\textbf{Video demo.} A video recording of the application that shows the real-time operation and user interaction with the biofeedback application can be accessed here: \href{https://osf.io/zhmtg/files/osfstorage/6866934e868a7d40f47ea837}{OSF video demo}.

\section*{ACKNOWLEDGMENT}
This work was supported by the Mobiliar Lab for Analytics at ETH Zürich. The authors gratefully acknowledge Delsys, Inc. for the loan of Trigno™ Avanti sensors and the NinaPro consortium for providing the EMG dataset used in our latency validation. AI Assistance was used to support the linguistic styling of this paper, for example, to help with paraphrasing or suggest refinements to author-drafted original content, rather than for suggesting new content.

\balance          
\bibliographystyle{IEEEtran}    % official IEEE style
\bibliography{IEEEabrv,references}  % first the abbreviation file, then yours

\end{document}